\documentclass[amsmath,amssymb,aps,showpacs,floatfix,pre,twocolumn ]{revtex4-1}
\usepackage{graphicx}
\usepackage{bm}
\usepackage{color}
\begin{document}

\title{Generalized model of blockage in particulate flow limited by channel carrying capacity.}
\author{C. Barr{\'e}$^1$, J. Talbot$^1$, P. Viot$^1$, \\
 L. Angelani$^2$, A. Gabrielli$^2$}
\affiliation{
$^1$Laboratoire de Physique Th\'eorique de la Mati\`ere Condens\'ee, UPMC, CNRS  UMR 7600, Sorbonne Universit{\'e}s,
 4, place Jussieu, 75252 Paris Cedex 05, France\\
$^2$Istituto dei Sistemi Complessi (ISC) - CNR, UOS ``Sapienza", Physics Department, University ``Sapienza" of
Rome, Piazzale Aldo Moro 2, 00185 - Rome, Italy
}
\begin{abstract}
 We investigate stochastic models of particles entering a channel with a random time distribution.
When the number of  particles  present in the channel exceeds a critical value $N$, a blockage occurs and the particle flux is definitively interrupted.
By introducing an integral representation of the $n$ particle survival probabilities, we obtain exact expressions for 
the survival probability, the distribution of the number of particles that pass before failure, 
the instantaneous flux of exiting particle and their time correlation.
We generalize previous results for $N=2$ to an arbitrary distribution of entry times and obtain new, exact solutions for $N=3$ for a Poisson distribution
and partial results for $N\ge 4$.
\end{abstract}
\date{\today}

\maketitle

\section{Introduction}

A stream of particles flowing through a channel may be slowed or blocked if the number of particles present 
exceeds the carrying capacity of the channel. This phenomenon is widespread and spans a 
range of lengthscales. Typical examples include vehicular and pedestrian traffic flow, 
filtration of particulate suspensions and the flow of macromolecules
through micro- or nano- channels. A specific example of the first category is a bridge that 
collapses if combined weight of the vehicular traffic exceeds 
a threshold. In filtration, experimental data of the fraction of grains retained by a filter mesh 
can be explained by assuming that clogging may occur when two or more grains are simultaneously present in the 
same vicinity of a mesh hole, even though isolated grains are small
enough to pass through the holes \cite{PhysRevLett.98.114502}. A biological example is provided by the bidirectional
traffic in narrow channels between the nuclear membrane and the cytoplasm\cite{Kapon2008}. 

The totally asymmetric simple exclusion effect process (TASEP) provides a theoretical approach to these phenomena. 
The TASEP is a  lattice model with a stochastic dynamics where particles hop randomly from site to site 
in one direction 
with the  condition that two particles cannot
occupy the same site at the same time \cite{Reuveni2012,Derrida1992}. 
At the two extremities of the finite lattice, particles are inserted and removed with two different rates. 
The model and its extensions provide quantitative descriptions 
of the circulation of cars and pedestrians\cite{Schiffmann2010,Moussa2012,Ezaki2012,Hilhorst2012a,Appert-Rolland2010,Appert-Rolland2011}. 
The  so-called
bridge models consider two TASEP processes
with oppositely directed 
flows, but allow exchange of particles
on the bridge\cite{Jelic2012,Grokinsky2007,Godreche1995,Evans1995a,Popkov2008}. At the microscopic level active motor protein transport on the 
cytoskeleton has been modeled by a TASEP \cite{Neri2013,Neri2013a}. 

Recently, some of the present authors  \cite{Gabrielli2013,TGV2015} introduced a class of continuous time and space stochastic models that are 
complementary to the TASEP approach. In these models particles
enter a passage at random times according to a given distribution. In the simplest concurrent model particles
move in the same direction and an isolated particle exits after a transit time $\tau$ 
but if $N=2$ particles are simultaneously present, blockage occurs. If the particle entries follow a homogeneous Poisson process all properties of 
interest, including the survival probability, mean survival time and the flux and distribution of exiting particles can be obtained analytically. 
The model has a connection to queuing theory in that it is a generalization of an M/D/1 queue, i.e. one where 
arrivals occur according to a Poisson process, service times are deterministic and with one server. 
This queue has many other applications including, for example, trunked mobile radio systems and airline hubs \cite{daiguji2004,barcelo1996,janssen2008}.

Opposing streams, where blockage is triggered by the simultaneous presence of two particles moving in different directions can be treated
within the same framework \cite{Gabrielli2013}. Inhomogeneous distributions of entering particles can be treated analytically \cite{BT2015}.
It is also possible to obtain exact solutions for when the blockage is of finite duration, rather than permanent \cite{BTV2013}.  
In this case, for a constant flux of incoming
particles the system reaches a steady-state with a finite flux of exiting particles that depends on the blockage time $\tau_b$.

The purpose of this article is to explore the properties of the concurrent flow models for any distribution of entry times and  when the threshold for blocking is $N>2$.
In addition to the applications described above, this generalized model may also be relevant 
for internet attacks, in particular  denial of service attacks (DoS) and a distributed denial of service attacks (DDoS) 
where criminals attempt to flood a network to prevent its operation\cite{kong2003,gao2011,bhunia2014}.

Unfortunately, the method used to solve the models for $N=2$ \cite{Gabrielli2013,BTV2013} applies only to a Poisson distribution and 
cannot be used even in this case for $N>2$. In section \ref{sec:cfm}, we develop a new approach providing formal 
exact expressions of the key quantities describing the kinetics of
the model. In section \ref{sec:cfm2}, as a first application, we recover the results of the model $N=2$  
that were first obtained by using a differential equation approach\cite{Gabrielli2013}.
In section \ref{sec:cfm3}, we present a complete solution when the entry time distribution is Poisson for $N=3$.
In section \ref{sec:cfm4} we consider the case of general $N$.
In section \ref{sec:cor} we investigate the time correlation for $N=2$ and $N=3$, 
and we further explore the model by studying the correlations between the
arrival times of the particles. We also explore the connection with the equilibrium properties of the hard rod fluid.

\section{Concurrent flow model }\label{sec:cfm}
\subsection{Definition}

We assume that at $t=0$ the channel of length $L$ is empty. The first particle enters at a time $t_0$ that is distributed according to a probability density
function $\psi(s)$. The entry of subsequent particles is characterized by the inter-particle time $t_i,i>0$ between the entry of particle $i$ and $i+1$. 
We assume
that the $t_i$ are distributed according to $\psi(s)$ and uncorrelated. 
The total elapsed time is then $t=t_0+\sum_{i=1}^{n-1}t_i+t'$ when $n$ particles have entered and $t'$ is
the time elapsed after the entry of the last particle. 

\begin{figure}[t]
\begin{center}
\includegraphics[width=8cm]{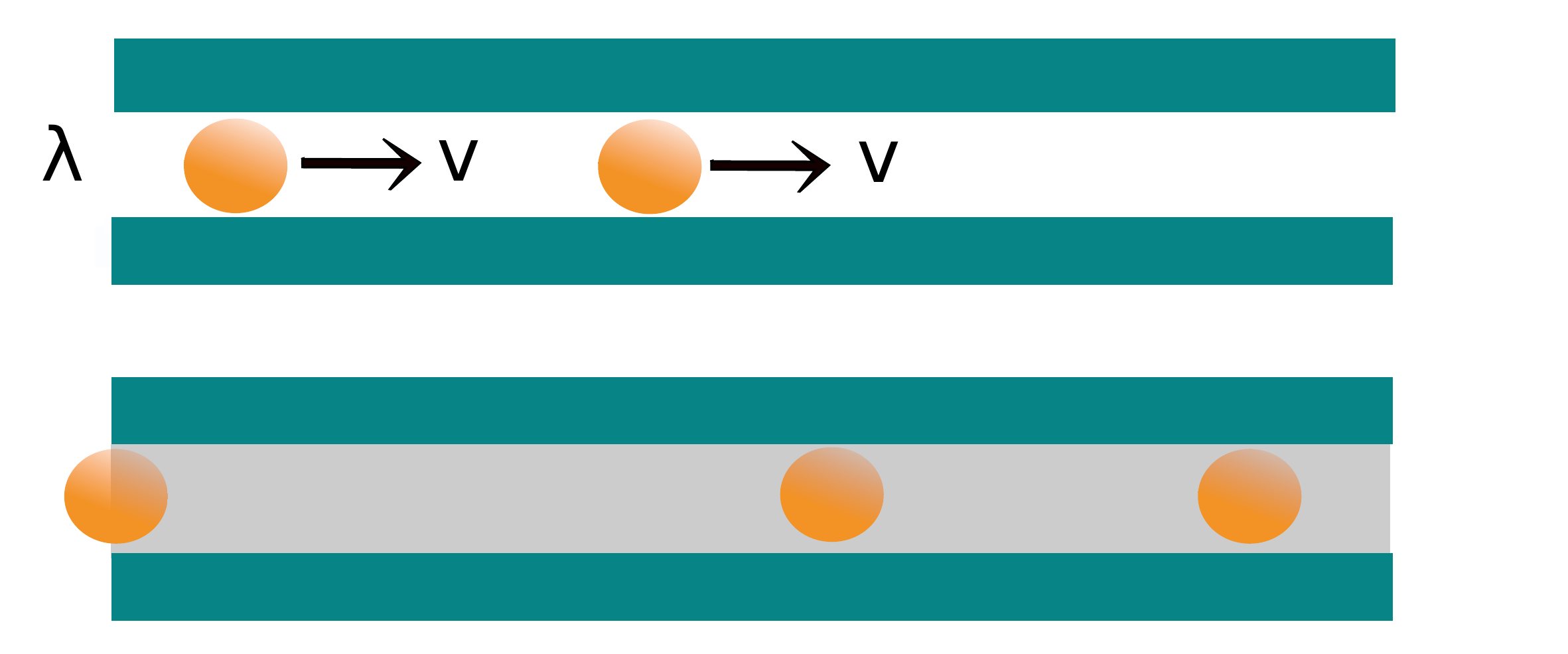}
\end{center}
\caption{Concurrent flow model $N=3$. Particles enter the left hand side of a channel of length
$L$ randomly. Top: two particles cross and exit the channel in a time $\tau$. Bottom: If a third particle enters while the two previous particles
are still in the channel, a blockage occurs instantaneously.}\label{fig:channel}
\end{figure}

If unimpeded
by the presence of another particle, a particle exits after
a transit time $\tau>0$. Blockage occurs when $N$ particles are present in the channel at the same time, which occurs if 
$t_i+t_{i+1}+\cdots+t_{i+N-2}<\tau$ (see Fig.\ref{fig:channel} for the case $N=3$). 
The model is non-Markovian as the state
of the system at time $t$  depends not only on the actual state but also on
the history of the system. 
The probability that no particle enters in the interval $[0,t]$ is $1-\psi_c(t)$ with $\psi_c(t)$ the cumulative distribution $\psi_c(t)=\int_0^t \psi(s)ds$.

The simplest case is a homogeneous Poisson process where the probability density function of particle times is $\psi(t)=\lambda e^{-\lambda t}$
where $\lambda$ is the rate (sometimes called the intensity).

\subsection{Quantities of interest} 

The key quantities describing the process are  the probability that the channel is active at time $t$, namely the survival probability, $p_s(t)$,
 the average blocking time $\langle t \rangle$ (where the bracket indicates an average over realizations of the process),  
 the number of particles that have exited  the channel at time $t$, 
 $\langle m(t)\rangle$,  and the instantaneous particle flux $j(t)$.

The survival probability can be expressed as the sum over all $n$-particle survival
probabilities $q(n, t)$, i.e. the joint probability
of surviving up to $t$ and that $n$ particles have entered
the passage during this time,
\begin{equation}\label{eq:ps}
p_s(t)=\sum_{n=0}^{\infty} q(n,t)
\end{equation}
For general $N$ and  $n> N-1$, $q(n,t)$ can be expressed as:
\begin{align}
q(n,t)=&\int_{0}^{\infty}\left[\prod_{i=0}^{n-1} dt_i\psi(t_i)\right] \int_{0}^{\infty}dt'(1-\psi_c(t'))\nonumber\\
&\left[\prod_{j=1}^{n-N+1}\theta\left(\sum_{m=0}^{N-2}t_{j+m}-\tau\right)\right]\delta\left(t-\sum_{i=0}^{n-1}t_i-t'\right) \label{eq:q1}
\end{align}
where $\theta(x)$ the Heaviside step function.
The first $n$ integrals correspond to the arrival of $n$ particles in the channel, with time intervals $t_i$, 
the integral over $t'$ imposes that no particle enters after particle $n$.
The Heaviside functions account for  the constraint that no consecutive sequence of $N$ particles can be simultaneously in the channel, i.e. in a time interval smaller than 
$\tau$ and
the $\delta$ function imposes that the observation time $t$ is equal to the sum of the time intervals $t_i$ plus $t_0$ and $t'$.

For $0\leq n\leq N-1$ there is no constraint on the particle time interval so the probability $q(n,t)$ is expressed as the joint probability of $n$ independent 
and identically distributed events
\begin{align}
q(n,t)=&\int_{0}^{\infty}\left[\prod_{i=0}^{n-1} dt_i\psi(t_i)\right] \int_{0}^{\infty}dt'(1-\psi_c(t'))\nonumber \\
&\delta\left(t-\sum_{i=0}^{n-1}t_i-t'\right) \label{eq:qq}
\end{align}
and \begin{equation}
 q(0,t)=1-\psi_c(t)
\end{equation}

Once the $q(n,t)$, and hence $p_s(t)$, are known we can obtain several useful quantities. 
The probability density function of the blocking time, $f(t)$ is simply related to $p_s(t)$
\begin{equation}
f(t)=- \frac {dp_s(t)}{dt} \label{eq:f1}
\end{equation}
Defining the Laplace transform as $\tilde{f}(u)=\int_0^\infty dt e^{-ut} f(t)$, one infers 
 \begin{equation}
 \tilde{f}(u)=1-u\tilde{p_s}(u)
\end{equation}  
The mean blocking time is given by  
\begin{equation}
\langle t \rangle=\int_0^{\infty} dt   t   f(t)= \tilde{p}_s(0)\label{eq:t}
\end{equation}

The instantaneous flux of particles exiting the channel 
can be obtained by noting that if a particle exits the channel at time $t$,
at most $N-1$ particles can enter the channel between $t$ and $t-\tau$ if no blockage is to occur. 
Since blockage is irreversible the flux tends to $0$ when the time increases, $j(\infty)=0$ for all value of $N$.
The total flux is given by the sum, 
\begin{equation}
j(t) = \sum_{n=1}^{\infty}j(n,t) \label{eq:jj}
\end{equation}
where $j(n,t)$ is the partial flux  where a particle exits the channel at time $t$ such that the channel is still open and $n$ particles have already entered,
for $n\geq N$
\begin{align}\label{eq:jn}
j(n,t) = &\int_0^{\infty} dt' (1-\psi_c(t'))\int_{0}^{\infty}\left[\prod_{i=0}^{n-1} dt_i\psi(t_i)\right]\nonumber\\
&\left[\prod_{j=1}^{n-N+1}\theta\left(\sum_{m=0}^{N-2}t_{j+m}-\tau\right)\right]\delta\left(t-\sum_{i=0}^{n-1}t_i-t'\right)\nonumber\\
&[\delta(t'-\tau)+\sum_{k=1}^{N-2} (\delta(t'+\sum_{w=1}^{k} t_{n-w} -\tau))],\;n\ge N
\end{align}

The  condition that a  particle exits at time $t$ is expressed in terms of $\delta$ functions. 
More specifically the exiting particle can be the last particle to enter, corresponding to the term $\delta(t'-\tau)$, or one of the other $N-1$ previously entering particles,  
corresponding to the sum over $\delta$ functions. 
For $n<N$ blocking is not possible, so Eq.(\ref{eq:jn}) is replaced by one without the Heaviside functions.  

Finally, the number of particles that have exited at time $t$ can be obtained by integrating over the particle flux
\begin{equation}
 \langle m(t)\rangle=\int_0^t dt' j(t')
\end{equation}

We can also obtain the distribution of particles exiting the channel. Let $h(m,t)$ denote the probability that blockage occurs in the interval $(0,t)$ 
and that $m$ particles 
have exited during this time. Its time evolution is  given by

\begin{align}
 \frac{dh(m,t)}{dt}=&\int_0^{\infty}\prod_{i=0}^{m+N-1}dt_i\psi(t_i)\prod_{j=1}^{m}\theta\left(\sum_{p=0}^{N-2}t_{j+p}-\tau\right)\nonumber\\
 &\theta(\tau-\sum_{p=1}^{N-1}t_{m+p})\delta(t-\sum_{i=0}^{m+N-1}t_i)\,m\geq1
\end{align}
The upper part of the right hand side corresponds to the event where $m+N$ particles have entered at time $t$,
and there was no blockage involving the first $m+N-1$ particles.
The second Heaviside function corresponds to the constraint that the last $N$ particles are blocked in the channel, 
with the $N+m^{th}$ particle entering at time $t$.

One can check that 
\begin{equation}
\left<m (t)\right>=\sum_{m=0}^{\infty}mh(m,t)=\int_0^{t}j(t')dt'
\end{equation}

We now consider the specific cases $N=2$ and $N=3$.

\section{$N=2$}\label{sec:cfm2}

Since each Heaviside function in Eq.(\ref{eq:q1}) depends on only one variable, the multiple integrals can be always calculated.  
Taking the Laplace transforms of Eq.(\ref{eq:q1}) and Eq.(\ref{eq:qq}), one obtains
\begin{equation}
\tilde q(n,u)=\tilde \psi(u) (\frac{1}{u}-\tilde \psi_c(u))\left[\int_{\tau}^{\infty}dt\,e^{-ut}\psi(t)\right]^{n-1}
\end{equation}
Using Eq.(\ref{eq:ps}) and $\tilde \psi_c(u)=\frac{\tilde{\psi(u)}}{u}$, we obtain the Laplace transform of the survival probability.
\begin{align}
\tilde p_s(u)&=\sum_{n=0}^{\infty}\tilde{q}(n,u)\nonumber\\
 &=\frac{1-\tilde{\psi}(u)}{u}\left(1+\frac{\tilde{\psi}(u)}{1-\int_{\tau}^{\infty}e^{-ut}\psi(t) dt}\right)
\end{align}

Therefore, the mean time of blockage is 
\begin{equation}
\langle t\rangle=\hat t \left[1+\frac{1}{\int_0^{\tau}\psi(t) dt}\right]\label{eq:t3}
\end{equation}
where $\hat t =\tilde{\psi}'(0)=\int_0^\infty dt\,t \psi(t)$ is mean inter particle time. 
To interpret Eq.(\ref{eq:t3}) we note that $ \psi_c(\tau)=\int_0^{\tau}\psi(t) dt$ gives the probability that 
two consecutive particles are separated by a time smaller than $\tau$. 

\begin{figure}[t]
\begin{center}
 \includegraphics[width=8.5cm]{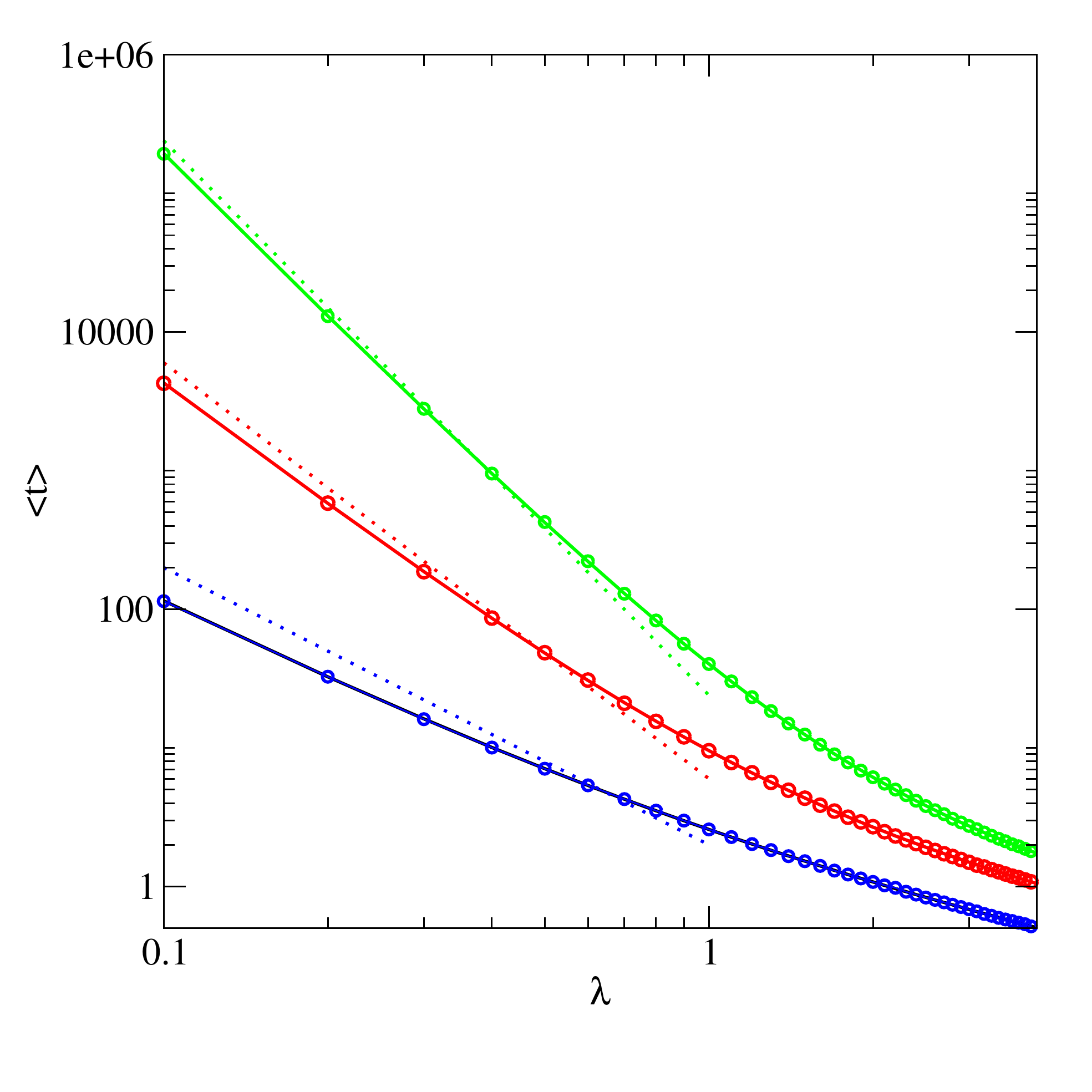}
\end{center}
\caption{Mean time of blocking $<t>$ as a function of the intensity,  $\lambda$, for a Gamma distribution 
for $\alpha=2,3,4$ (from bottom to top), 
from numerical simulation (circles) and Eq.(\ref{eq:t3})  (full lines). Dotted lines correspond to the asymptotic behavior, Eq.(\ref{eq:tempsmoyengamma}). }
\label{fig:1bis}
\end{figure}

For a Gamma distribution, $\psi(t)=\lambda^\alpha t^{\alpha-1}e^{-\lambda t}/\Gamma(\alpha)$ where $\alpha$ is a shape parameter, the mean time of blocking is equal to
\begin{equation}
 \langle t\rangle=\frac{\alpha}{\lambda}\left(1+\frac{\Gamma(\alpha)}{\Gamma(\alpha)-\gamma(\tau,\alpha)}\right)\label{eq:tempsmoyengamma}
\end{equation}
where $\Gamma(\alpha)$ and $\gamma(\alpha,x)$ are the Gamma and incomplete Gamma functions, respectively.
When $\lambda\tau< 1$, one obtains 
\begin{equation}
 \langle t\rangle=\frac{1}{\lambda}\frac{\alpha!}{(\lambda\tau)^\alpha}\label{eq:tempsmoyengamma2}
\end{equation}
Figure \ref{fig:1bis} shows $<t>$ versus  $\lambda\tau$  for  $\alpha=2,3,4$. One observes an excellent agreement between simulation data (circles) and the exact formula, 
Eq.(\ref{eq:tempsmoyengamma}). As expected, the mean time $<t>$ diverges when $\lambda$ goes to zero. The asymptotic behavior, Eq.(\ref{eq:tempsmoyengamma2}) provides a good approximation
of  simulation data when $\lambda\tau<1$.

By taking  $\alpha=1$ in the Gamma distribution, which corresponds to a homogeneous Poisson process, the mean time of blockage is given by  
\begin{equation}
\langle t \rangle=\frac{2-e^{-\lambda \tau}}{\lambda(1-e^{-\lambda \tau})}\label{tempsn2}
\end{equation}
a result previously obtained by using a master equation for  the time evolution of the $q(n,t)$ \cite{Gabrielli2013}.

The mean flux $j(t)$ can be obtained by using  Eqs.(\ref{eq:jj},\ref{eq:jn})
\begin{align}
j(t) = &\sum_{n=1}^{\infty}\int_0^{\infty} dt'  (1-\psi_c(t'))\int_{0}^{\infty}\left[\prod_{i=0}^{n-1} dt_i\psi(t_i)\right]\nonumber\\
&\left[\prod_{j=1}^{n-1}\theta\left(t_j-\tau\right)\right]\delta\left(t-\sum_{i=0}^{n-1} t_i-t' \right)[\delta(t'-\tau)]
\end{align}

The multiple integral can be factorized and the flux is given by :
\begin{align}
 j(t)=&\sum_{n=1}^{\infty}\int_0^{\infty}dt_0 \psi(t_0)\int_0^{\infty}dt'(1-\psi_c(t'))\left[\int_{\tau}^{\infty}dt\psi(t)\right]^{n-1}\nonumber \\
 &\delta\left(t-\sum_{i=0}^{n-1}t_i\right)[\delta(t'-\tau)]
\end{align}

In Laplace space, the summation over $n$ can be performed and $\tilde{j}(u)$ is given by
\begin{align}\label{eq:jgeneral}
\tilde{j}(u) =\frac{(1-\psi_c(\tau))e^{-u\tau} \tilde \psi(u)}{1-\int_{\tau}^{\infty}e^{-ut}\psi(t) dt}
\end{align}
With a Poisson distribution $\psi(t)=\lambda e^{-\lambda t}$, we have
\begin{equation}
\tilde{j}(u) =\frac{\lambda e^{-(u+\lambda)\tau} }{u+\lambda(1-e^{-(u+\lambda)\tau})}
\end{equation}
By taking the inverse Laplace transform, the mean flux $j(t)$ can be expressed as a series
\begin{equation}
j(t)=\lambda e^{-\lambda t} \sum_{n=1}^\infty 
\left[ \frac{1}{n!}(\lambda (t-(n+1)\tau))^{n}\theta(\lambda (t-(n+1)\tau)) \right]
\end{equation}
as obtained previously by using a master equation approach \cite{TGV2015}. 

No particle exits the channel  between $0$ and $\tau$; indeed, the flux is obviously equal to $0$ in this interval  and  rises instantaneously to a maximum,
$j_{\max}=\lambda e^{-\lambda \tau}$ which itself is maximum when $\lambda =\frac{1}{\tau}$, and then decreases to $0$. 

For a Gamma distribution  with an integer value of $\alpha$, 
the Laplace transform of the flux can be obtained explicitly, but increasing $\alpha$ it rapidly leads to  lengthy expressions.
\begin{figure}[t]
\begin{center}
\includegraphics[width=8cm]{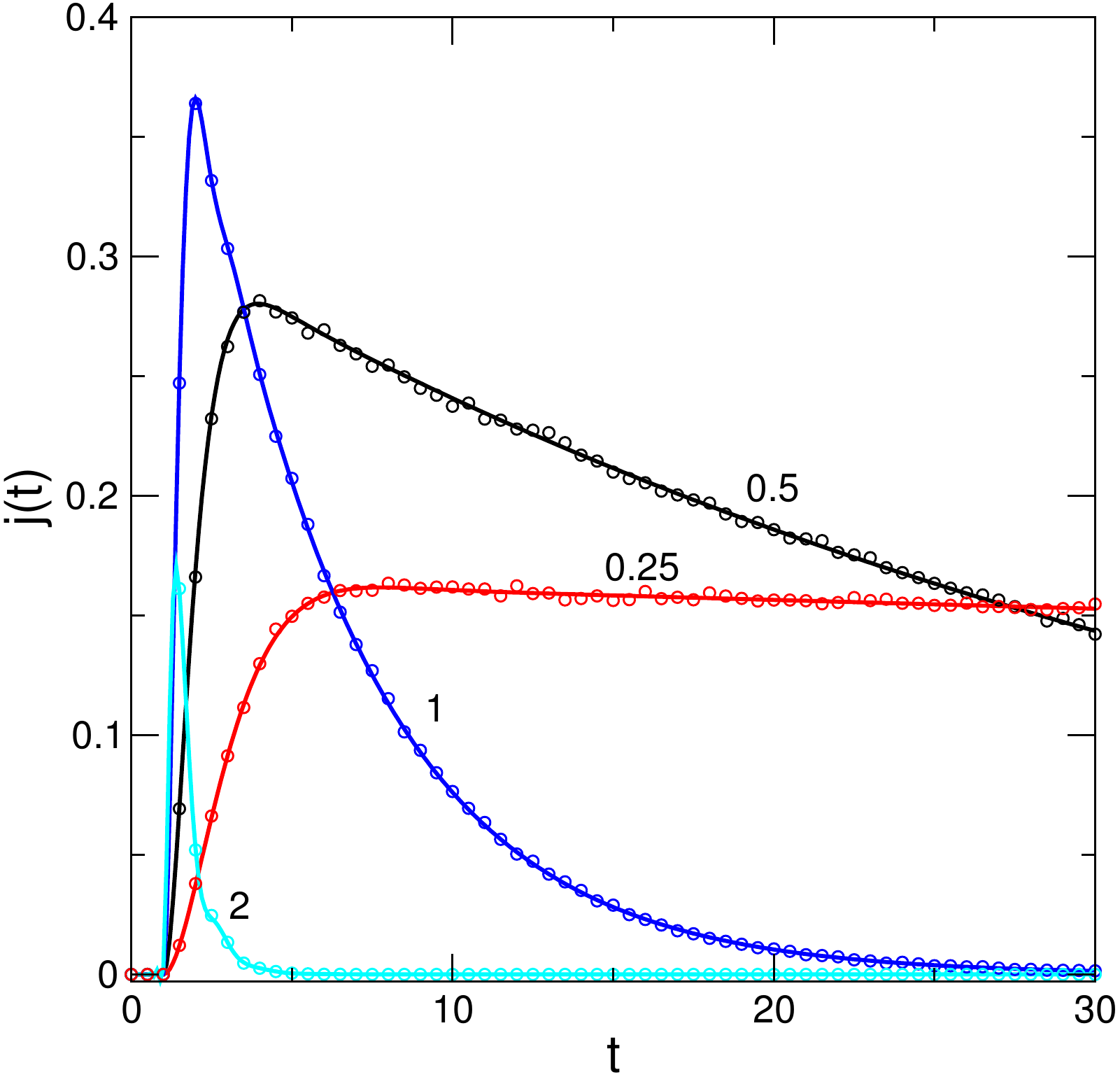}
\end{center}
\caption{Mean flux $j(t)$ as a function of time for a Gamma distribution with $\alpha=3$ and for different values 
of $\lambda$ ($\tau=1$). The solid lines show the exact expression, Eq.(\ref{eq:jgeneral}) and the circles show simulation results.}
\label{fig:fluxgamma}
\end{figure}

Figure \ref{fig:fluxgamma} displays the time evolution of the mean flux  $j(t)$ for different values of 
$\lambda$ and $\alpha=3$. In all cases, the flux becomes nonzero for $t>\tau$, corresponding to the exit of a first particle. 
For $\lambda\tau\geq 1$, $j(t)$ displays a strong maximum
at a a time $t_m$ slightly larger than $\tau$ and decays to $0$. For $\lambda\tau=0.5$, the  maximum of the flux is shifted to a time $t_m\simeq 3\tau$ and the typical decay time   is around $100\tau$.
For $\lambda\tau=0.25$, $j(t)$ increases up to a quasi-plateau and the typical decay time is larger than $1000\tau$, which corresponds to a physical situation where a large number of particles exit the channel
before the definitive clogging.
Note that for a given value of $\lambda$ the flux
is much larger than for a Poisson distribution. 
However, it approaches zero for sufficiently long times with a characteristic time equal to the mean blocking time.

We also consider the probability, $h(m,t)$, that blockage occurs in the interval $(0,t)$ and that during this time $m$ particles exit the channel.
The time evolution of this function is given by
\begin{align}
 \frac{dh(m,t)}{dt}=&\int_0^{\infty}\prod_{i=0}^{m+1}dt_i\psi(t_i)\prod_{j=1}^{m}\theta\left(t_{j}-\tau\right)\nonumber \\
 &\theta(\tau-t_{m+1})\delta(t-\sum_{i=0}^{m+1}t_i)
\end{align}
Two particles have to be in the channel for the system to block, so the interval between $2$ consecutive 
particles has to be less than $\tau$ (the $\theta$ function). 
The previously entering particles exited the channel without blockage.

Taking the Laplace transform  we obtain  for $m\geq 0$

\begin{align}
 \tilde{h}(m,u)=\frac{\tilde{\psi}(u)}{u}\int_0^{\tau}\psi(t')e^{-ut'}dt'
 \left[\int_\tau^{\infty}dt\psi(t)e^{-ut}\right]^{m}\label{eq:hm}
\end{align}
The probability that the channel is blocked can be expressed as the sum over partial probabilities $h(n,t)$, namely $h(t)=\sum_{m=0}^{\infty}h(m,t)$. 
By using Eq.(\ref{eq:hm}), one infers 
$\lim\limits_{t \rightarrow \infty}h(t)=\lim\limits_{u \rightarrow 0}u h(u)=1$, as because blockage is certain to occur, 
a result valid for any distribution $\psi(t)$. 
Finally, we note the following sum rule, $\sum_{n\geq 0}(q_s(n,t)+h(n,t))=1$ - all configurations of the process are either
blocked or unblocked.

For the Poisson process, an explicit expression can be obtained
\begin{align}
 \tilde{h}(m,u)=\frac{\lambda^{m+2}}{u(\lambda+u)^{m+2}}\left[1-e^{-(\lambda+u)\tau}\right]e^{-(\lambda+u)m \tau}
\end{align}
Performing the Laplace inversion we obtain $h(m,t)$  as obtained previously \cite{TGV2015}.
As expected, $h(m, t)$ is equal to zero for $t < m\tau$ corresponding to the minimum time necessary for $m$ particles to exit the channel.
For the Gamma distribution with $\alpha=2$ we obtain 
\begin{align}
 \tilde{h}(m,u)=&\frac{1}{u(u+\lambda)^{2(m+2)}}(e^{-(u+\lambda)\tau m}\lambda^{2(m+2)}\nonumber\\
 &(1+(u+\lambda)\tau)^m(1-e^{-\tau(u+\lambda)}(1+(u+\lambda)\tau)))\label{eq:hgamma}
\end{align}

\begin{figure}[t]
\begin{center}
\includegraphics[width=8cm]{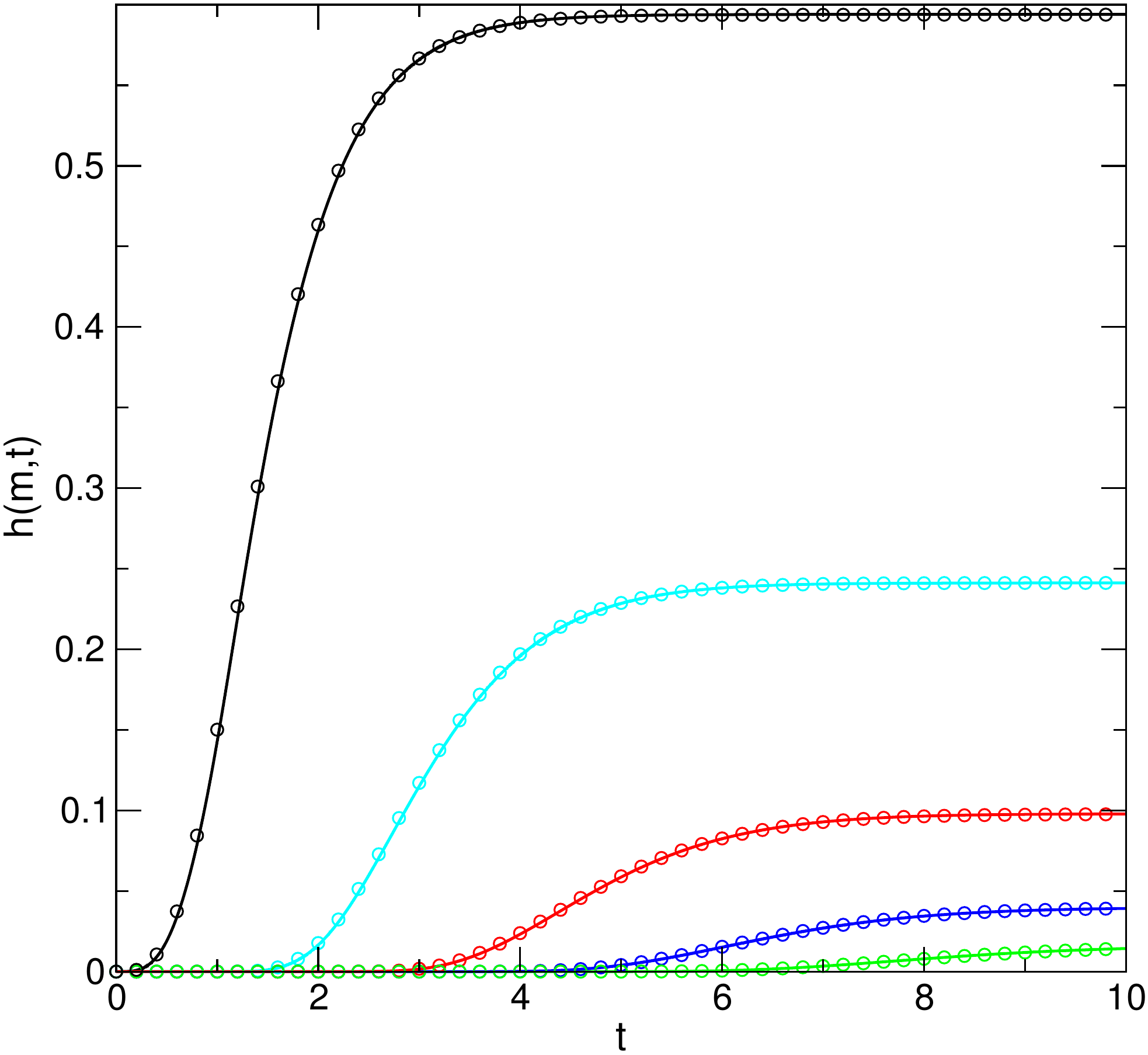}
\end{center}
\caption{The probability $h(m,t)$ as a function of time for a Gamma distribution with $\alpha=2$, $\lambda=2$ and  $m=0,1,2$ (from top to bottom).
The full curves show the exact expression, Eq. (\ref{eq:hgamma}) and circles show simulation results.}
\label{fig:gammahn}
\end{figure}

For the Gamma distribution,  we plot in 
Fig. \ref{fig:gammahn}  the time evolution of $h(m,t)$ as a function of time with  $m=0,1,2$ for  $\alpha=2$ and $\lambda=2$.
As expected,  $h(m,t)=0$ for $t<m\tau$ , which can be explained by the fact that
the minimum  time for having a configuration where $m$ particles exit the channel must be at least larger than $m\tau$. 
Similarly, the transient time associated with $h(m,t)$ increases with $m$, and corresponds to rare events 
when $m$ increases.

\section{$N=3$}\label{sec:cfm3}
For the first three partial probabilities, there is no constraint and one easily obtains that 
$q(0,t)=1-\psi_c(t)$,  and for $i=1,2$ the probabilities are given in terms of the Laplace transforms 
$q(i,u)=\left(\frac{1-\tilde{\psi}(t)}{u}\right)\tilde{\psi}(u)^i$. For a Poisson process, one recovers that
$q(0,t)= e^{-\lambda t}$, $q(1,t)=\lambda t e^{-\lambda t}$ and $q(2,t)=\frac{(\lambda t)^2}{2}e^{-\lambda t}$.  
For $n>2$, Eq.(\ref{eq:q1}) becomes 
\begin{align}\label{eq:qn3}
q(n,t)=&\int_{0}^{\infty}dt'(1-\psi_c(t')) \int_0^{\infty}\prod_{i=0}^{n-1} dt_i \psi(t_i) \nonumber\\
&\times \prod_{j=1}^{n-2} \theta(t_{j}+t_{j+1}-\tau) \delta(t-\sum_{i=0}^{n-1}t_{i}-t')
\end{align}
The constraint, imposed by the $\theta$ function, requires that the sum of two consecutive time intervals be less than $\tau$.
Taking  the Laplace transform of Eq.(\ref{eq:qn3}), one obtains 
 \begin{align}\label{eq:qn}
 \tilde{q}(n,u)=& \frac{\tilde{\psi}(u)(1-\tilde \psi(u))}{u}\int_{0}^{\infty}dt \psi(t)e^{- ut}r(n-1,t,u)
 \end{align}
where  the auxiliary function $r(n-1,t,u)$ is given by
  \begin{align}
 r(n-1,t,u)=\int_0^{\infty}\prod_{i=1}^{n-2}dt_i \psi(t_i) e^{-ut_i}\prod_{j=1}^{n-2}\theta(t_{j}+t_{j+1}-\tau) 
  \end{align}
  where $t_{n-1}=t$. A recurrence relation can be written for $r(n,t,u)$
  \begin{equation}\label{eq:recu}
 r(n,t,u)=\int_{\max(\tau-t,0)}^{\infty} dt' \psi(t')e^{-ut'} r(n-1,t',u)
  \end{equation}
  with $r(1,t,u)=1$.

Let us introduce  the generating function  $G_r(z,t,u)$ defined as
 \begin{equation}\label{ei}
  G_r(z,t,u) = \sum _{n=1} z^{n-1} r(n,t,u) 
 \end{equation}  
Multiplying Eq.(\ref{eq:recu}) by $z^{n-1}$ and summing over $n$, one obtains that
\begin{equation}\label{eq:ei}
 G_r(z,t,u)=1+z\int_{\max(\tau-t,0)}^{\infty} dt' \psi(t')e^{-ut'} G_r(z,t',u)
\end{equation}

For $t>\tau$  $G_r(z,t,u)$ is constant, i.e. $G_r(z,t,u)=G_r(z, \tau,u)$. 
For $t<\tau$, it is convenient to express the time evolution of $G_r(z,t,u)$ as follows:
taking the first two partial derivatives of $G(z,t,u)$ 
with respect to $t$, one obtains the ordinary differential equation 
\begin{align}
\frac{\partial^2 G_r(z,t,u)}{\partial t^2}&=\left(-\frac{\dot{\psi}(\tau-t)}{\psi(\tau-t)} + u\right)\frac{\partial G_r(z,t,u)}{\partial t}
\nonumber\\&-z^2 \psi(\tau-t)\psi(t)e^{-u\tau}G_r(z,t)\label{eq:ode1}
\end{align}
By using Eq.(\ref{eq:ei}), the differential is supplemented by two  boundary conditions
\begin{align}
\left\{ 
\begin{array}{ll}
G_r(z,0,u)&=1 + z G_r(z,\tau,u)\int_{\tau}^ {\infty}dt' \psi(t')e^{-ut'} \\
\left.\frac{\partial G_r(z,t,u)}{\partial t}\right|_{t=\tau}&=z \psi(0) G_r(z,0,u)
\end{array}
\right.  \label{eq:ci1}
\end{align}

Eq.(\ref{eq:ode1}) cannot be solved analytically in general but for a Poisson distribution it becomes
\begin{align}
\frac{\partial^2 G_r(z,t,u)}{\partial t^2}&=\left(\lambda + u\right)\frac{\partial G_r(z,t,u)}{\partial t}\nonumber \\
&-(z\lambda)^2e^{-(u+\lambda)\tau}G_r(z,t,u)\label{eq:ode2}
\end{align}
with the boundary condition given by Eq.(\ref{eq:ci1}) with $\psi (t)=e^{-\lambda t}$.

The solutions of the characteristic equation of Eq.(\ref{eq:ode2}) are 
\begin{equation}
 s_{1,2}(z,u)= \frac{(\lambda+u)\pm \sqrt{(\lambda+u)^{2}-4(z\lambda)^{2}e^{-(\lambda+u)\tau}}}{2}
\end{equation} 
and the generating function is given by $ G_r(z,t,u)=A(z,u)e^{s_{1}(z,u)t}+B(z,u)e^{s_{2}(z,u)t}$ where $A(z,u)$ and $B(z,u)$ are determined
by Eq.(\ref{eq:ci1}) adapted to a Poisson process. 

For $n=0,1,2$ the partial probabilities $q(n,t)$ correspond to those of a Poisson process. For $n>2$, by using the generating function $G_r(z,u)$, the Laplace transform
of $q(n,t)$ is given by
\begin{align}
  \tilde{q}(n,u)=\frac{\lambda }{(\lambda+u)^2} \int_{0}^{\infty}dt \lambda  e^{-(\lambda+u) t}
  \left. \frac{\partial^{n-2} G_r(z,t,u)}{\partial z^{n-2}} \right|_{z=0}
\end{align}
After some calculation, one obtains
\begin{align} 
\begin{array}{ll}
 \displaystyle q(3,t)&= \theta(t-\tau) \lambda^3 e^{-\lambda t}\left[\frac{1}{2} \tau (t-\tau)^2 +\frac{1}{6}(t-\tau)^3\right]\\
 \displaystyle  q(4,t)&=   \lambda^4 e^{-\lambda t}\left[\theta(t-\tau)\frac{(t-\tau)^4}{12}   -\theta(t-2\tau)  \frac{(t-2\tau)^4}{24}\right]
\end{array}  
\end{align}

By using Eqs.(\ref{eq:qn}) and (\ref{ei}), the Laplace transform of the survival probability $p_s(t)$ is
\begin{align}
 \tilde{p}_s(u)=&\tilde{q}(0,u)+\tilde{q}(1,u)+\nonumber\\
 &\frac{\psi(u)(1-\tilde \psi(u))}{u}\int_{0}^{\infty}dt \psi(t)e^{- ut}G_r(1,t,u)
\end{align}

By inserting the solution of Eq.(\ref{eq:ode1}), the Laplace transform of the survival probability is given by 
\begin{align}
  \tilde{p}_s(u)&=\frac{\lambda}{(\lambda+u)^2}\left[1+\frac{u}{\lambda}+
  A(1,u)\left[1+\frac{\lambda}{s_2}(1- e^{-s_2 \tau})\right]\right.
 \nonumber\\
  &+\left. B(1,u)\left[1+\frac{\lambda}{s_1}(1- e^{- s_1\tau})\right]\right]
\end{align}
where
\begin{align}
  A(1,u)=&\frac{\lambda e^{s_2\tau} (s_2-\lambda)(s_1+s_2)}{\Delta}
\nonumber\\
 B(1,u)=&\frac{\lambda e^{s_1\tau} (s_1-\lambda)(s_1+s_2)}{\Delta}\label{eq:a1b1}
\end{align}
with \begin{equation}
 \Delta= e^{(s_1+s_2\tau)}s_1s_2(s_1-s_2)+\lambda(s_2^2 e^{s_2\tau}-s_1^2 e^{s_1\tau})
\end{equation}

\begin{figure}[t]
\begin{center}
\includegraphics[width=8cm]{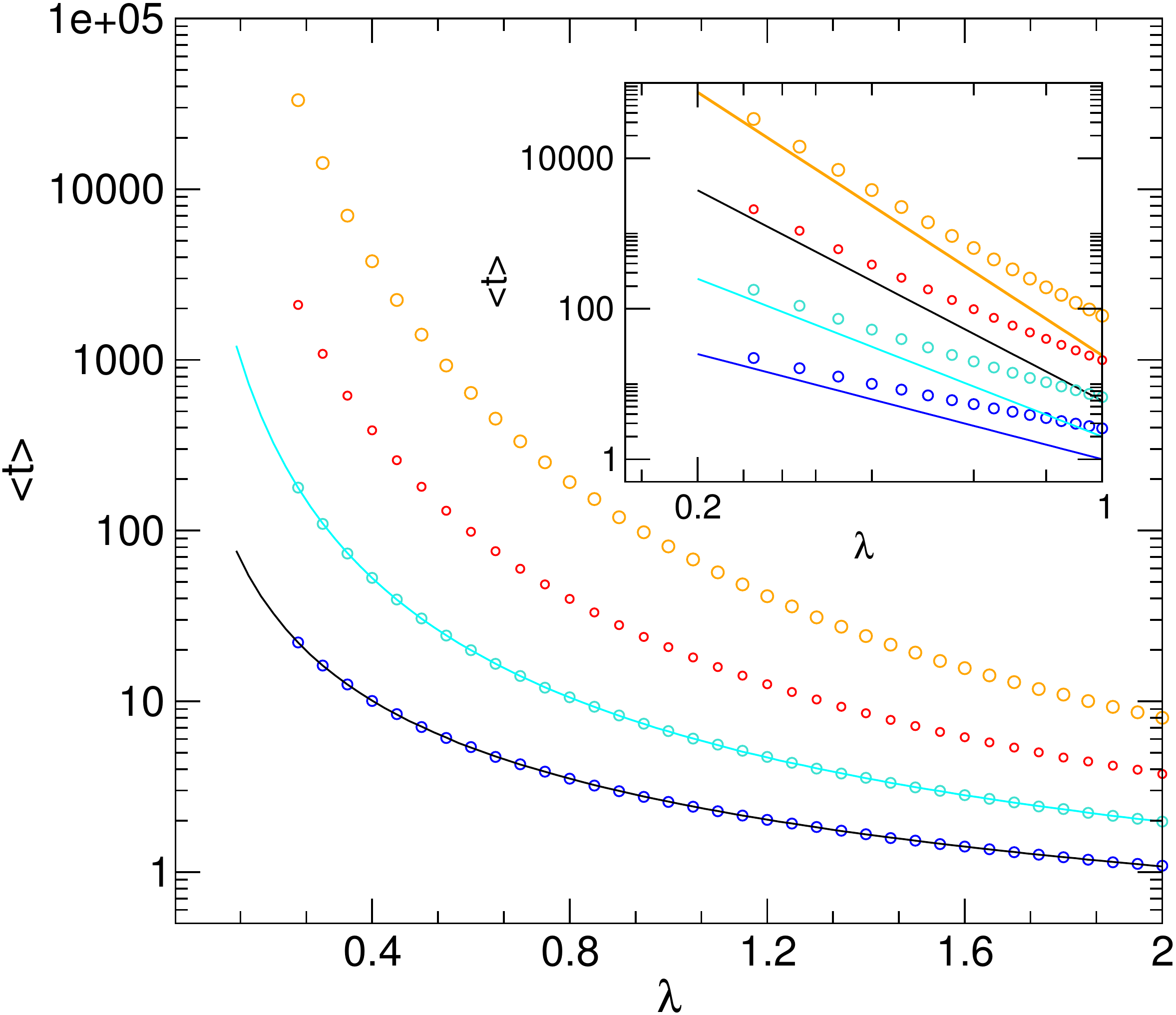}
\end{center}
\caption{Mean time of blocking as a function of the intensity  $\lambda \tau$ for $N=5,4,3,2$, top-to-bottom,  
from numerical simulation (circles) and Eq.(\ref{tempsn2}) $N=2$ and Eqs.(\ref{temps},\ref{temps2}) $N=3$ (full curves) for a Poisson distribution. 
The inset compares the asymptotic formula, Eq.(\ref{eq:tN}), with simulation results.}
\label{fig:tbarN}
\end{figure}

From the generating function, one can also obtain global quantities, like the mean blocking time $\langle t \rangle=\tilde{p}_s(0)$

Let $g=\sqrt{|1-4e^{-\lambda\tau}|}$ and $\nu=\frac{\lambda \tau}{2}$ then, after some calculation, one obtains for $\lambda\tau>2\ln(2)$ 
\begin{equation}
\lambda \langle t\rangle= \frac{2e^{\nu}\sinh(\nu)+g e^{\lambda\tau}}{-g-2\sinh(\nu)
e^{-\nu}+e^{\nu}\left(\sinh(\nu)+g\cosh(\nu)\right)}+1\label{temps}
\end{equation}
and for  $\lambda\tau<2\ln(2)$
\begin{equation}
\lambda \langle t\rangle= \frac{2e^{\nu}\sin(\nu)+g e^{\lambda\tau}}{-g-2\sin(\nu)
e^{-\nu}+e^{\nu}\left(\sin(\nu)+g\cos(\nu)\right)}+1\label{temps2}
\end{equation}

Fig. \ref{fig:tbarN} shows  the mean blocking time $\langle t \rangle$ of  the models with $N=2,3,4,5,6,7$ for 
a Poisson distribution obtained by simulation and for $N=2,3$
by using the analytic expressions. We observe a perfect agreement between simulation data and  exact expressions 
for $N=2$ Eq.(\ref{tempsn2}) and $N=3$ Eq.(\ref{temps},\ref{temps2}). 
More generally, one observes a divergence of 
the mean blocking time  as $\lambda\tau$ goes to $0$ and indeed
performing a first-order expansion of Eq.(\ref{temps2}) in $\lambda\tau$ gives
\begin{equation}\
\langle t\rangle\simeq \frac{2\tau}{(\lambda\tau)^3}
\end{equation}

The mean flux $j(t)$ can be also obtained by using Eq.(\ref{eq:jn}) and the auxiliary functions $r(n,t,u)$ and it comes for the Laplace transform $\tilde{j}(n,u)$ (for $n\geq 1$)
\begin{align}
 \tilde{j}(n,u)=&e^{-u\tau}\tilde{\psi}(u)\left((1-\psi_c(\tau))\int_0^{\infty} dt e^{-ut}\psi(t)r(n-1,t,u)\right.\nonumber\\
 &+\left.\int_0^\tau dt \psi(t)(1-\psi_c(\tau-t))r(n-1,t,u)\right)\label{eq:fluxj3}
\end{align}

By summing over $n$ (accounting for the boundary terms $\tilde{j}(1,u)$ and $\tilde{j}(2,u)$, the Laplace transform $\tilde{j}(u)$ is expressed as

\begin{align}
 \tilde{j}(u)=&e^{-u\tau}\tilde{\psi}(u)(1-\psi_c(\tau))\int_0^{\infty} dt e^{-ut}\psi(t) G_r(1,t,u)\nonumber\\
 &+e^{-u\tau}\tilde{\psi}(u)\int_0^\tau dt \psi(t)(1-\psi_c(\tau-t))G_r(1,t,u)\nonumber\\
&+\tilde{j}(1,u)
\label{eq:fluxj3}
\end{align}
By using Eq.(\ref{eq:ci1}) and the expression of the generating function $G_r(1,t,u)$, the Laplace transform of the flux can be expressed as
\begin{align}
  \tilde{j}(u)=&\frac{\lambda e^{-(u+\lambda)\tau} }{\lambda+u}\left[A(1,u)\left(e^{s_1\tau}\left(1+\frac{\lambda}{s_1}\right)-\frac{\lambda}{s_1}\right)
\right.\nonumber\\
&+\left.B(1,u)\left(e^{s_2\tau}\left(1+\frac{\lambda}{s_2}\right)-\frac{\lambda}{s_2}\right)\right]\label{eq:fluxj3bis}
\end{align}
where $A(1,u)$ and $B(1,u)$ are given by Eq.(\ref{eq:a1b1}).

Because  the right-hand-side of Eq.(\ref{eq:fluxj3bis}) can be factorized by $e^{-u\tau}$, it implies that $j(t)=0$  
for $t<\tau$, which corresponds to the minimum time for a particle to exit the channel.

\begin{figure}[t]
\begin{center}
\includegraphics[width=8cm]{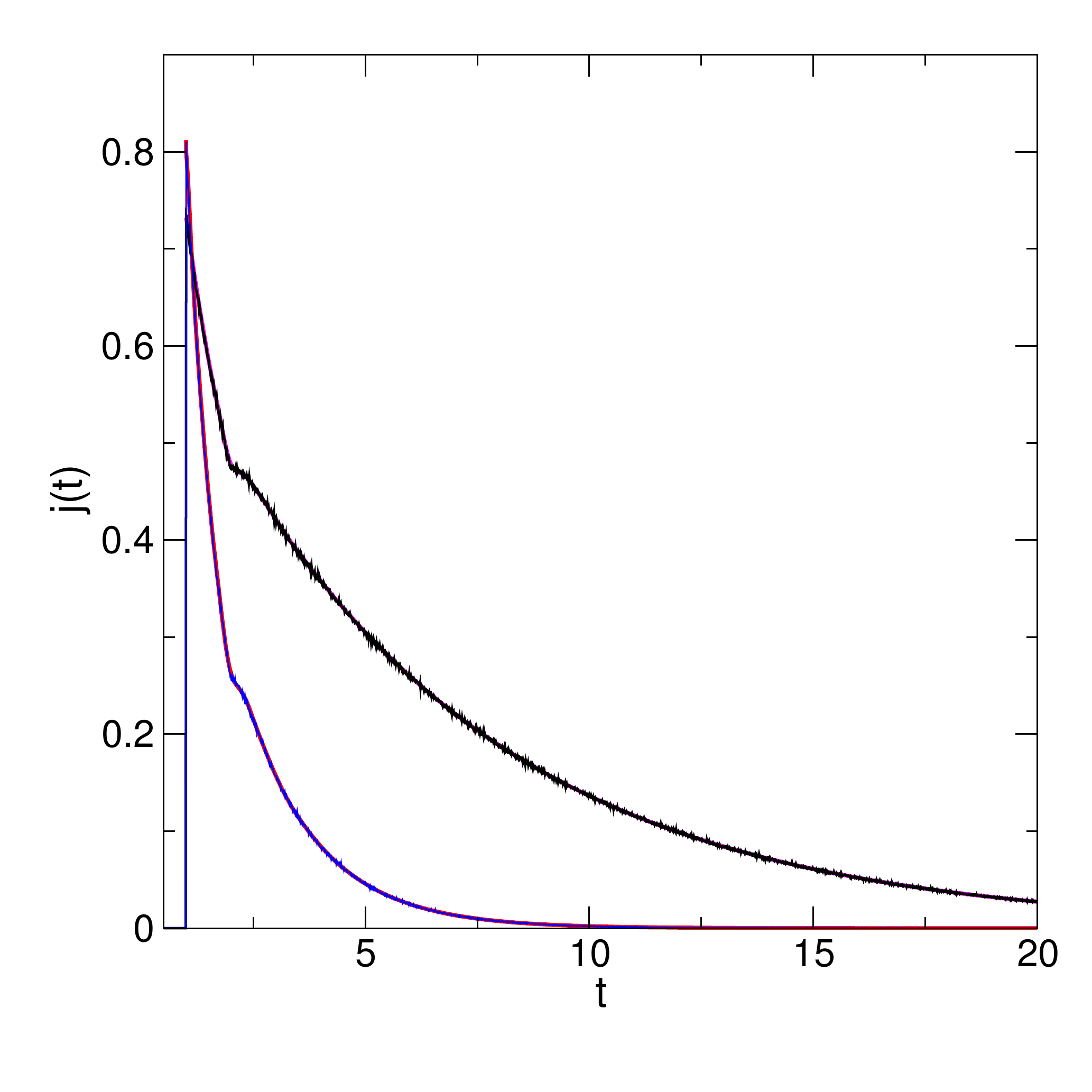}
\end{center}
\caption{Mean flux $j(t)$ as a function of time for a Poisson  distribution   for  
 $\lambda=1,2$ ($\tau=1$). The solid lines show the exact expression, (inverse Laplace transform of Eq.(\ref{eq:fluxj3bis})) accurately
matching    simulation results (wavy lines).}
\label{fig:fluxgamma3}
\end{figure}

The mean flux $j(t)$ is plotted as as function of time for  $\lambda=1,2$ with a Poisson distribution for $\lambda=1,2$ (Fig. \ref{fig:fluxgamma3}).
A discontinuity appears at $t=\tau$ where the flux is maximum $j(\tau)=\lambda$. At $t=\tau$, the flux is given by
\begin{equation}
 j(\tau)=\lambda(1+\lambda\tau)e^{-\lambda\tau}
\end{equation} which 
corresponds to events where a particle exits between $t$ and $t+dt$ such that $0$ or $1$ particle is still in the channel. 
The flux decay exhibits  a visible cusp at $t=2\tau$ which corresponds to the non analytical structure of the solution.
At long times, the flux decays to $0$, with a typical time which becomes larger when $\lambda$ decreases.

The joint probability $h(m,t)$ can also be obtained with the function $r(n,t,u)$.
For $m\ge1$ its time evolution is given by 
\begin{align}
 \frac{dh(m,t)}{dt}=&\int_0^{\infty}\prod_{i=0}^{m+2}dt_i\psi(t_i)\prod_{j=1}^{m}\theta\left(t_{j}+t_{j+1}-\tau\right)\nonumber\\
 &\theta(\tau-t_{m+1}-t_{m+2})\delta(t-\sum_{i=0}^{m+2}t_i) 
\end{align}

Taking the Laplace transform gives
\begin{align}
 \tilde{h}(m,u)=&\frac{\tilde{\psi}(u)}{u}\int_0^{\infty} \prod_{i=1}^{m+2}dt_i\psi(t_i)e^{-ut_i}\nonumber\\
 &\prod_{j=1}^{m}\theta(t_{j}+t_{j+1}-\tau)\theta(\tau-t_{m+1}-t_{m+2})
\end{align}

that can be expressed using the function $r(n,t,u)$ as 
\begin{align}
 \tilde{h}(m,u)=&\frac{\tilde{\psi}(u)}{u}\int_0^\tau dt\psi(t)e^{-ut}\nonumber\\
 &\int_0^{\tau-t}dt'\psi(t')e^{-ut'}r(m+1,t',u)
\end{align}

\begin{figure}[t]
\begin{center}
\includegraphics[width=8cm]{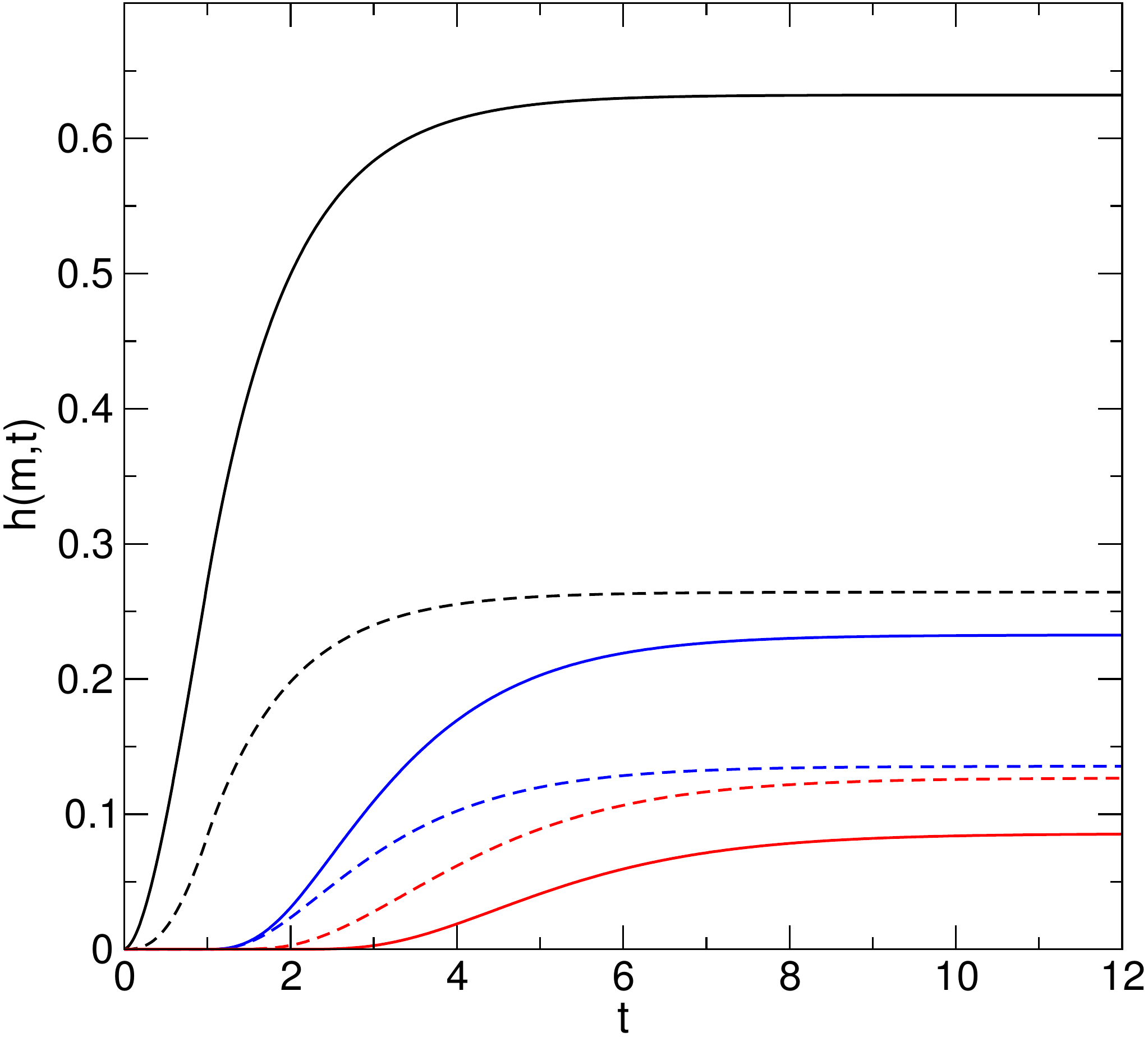}
\end{center}
\caption{Probability distributions $h(m,t)$ versus time $t$ for a Poisson distribution, for $m=0,1,2$ (from top to bottom) and $\lambda=1$. The solid lines correspond to  the model with $N=2$  and dashed lines correspond
to the  model with $N=3$.}
\label{fig:hpoisson}
\end{figure}

Figure \ref{fig:hpoisson} shows  the time evolution of the 
probability distributions $h(m,t)$ for a Poisson distribution with $\lambda=1$.
The probability that zero or one particle ($m=0,1$) particle exits is smaller for $N=2$ than for $N=3$. 
For $m\ge 2$ the order reverses (e.g. for $m=5$, case shown). 
This is because, for a given value of $\lambda \tau$, more particles exit before blockage as $N$ increases.

\section{$N\ge4$}\label{sec:cfm4}
We have seen that for $N=3$ the product of Heaviside functions in Eq.(\ref{eq:q1}) leads to a 
simple recurrence relation Eq.(\ref{eq:recu}). 
For $N\geq 4$ the task is much more difficult
because one needs to introduce  auxiliary functions that depend on $N-2$ time variables. These functions are related by an integral 
equation that cannot be converted to an ordinary differential 
equation. We therefore propose an approximate treatment of the dynamics. 
For the model where the blockage occurs when $N$ particles enter the channel between $t-\tau$ and $t$, 
the first $N-2$ partial probabilities $q(i,t)$ obey differential equations identical to those of a Poisson process
\begin{equation}
 \frac{d q(0,t)}{dt}=-\lambda q(0,t)
\end{equation}
and
\begin{equation}
\frac{d q(n,t)}{dt}=-\lambda q(n,t)+\lambda q(n-1,t),\;1\le n \le N-1
\end{equation}
For $n>N-1$,  the non-Markovian constraint applies, but for  $n=N$, the time evolution is simply given by 
\begin{equation}
\frac{d q(N,t)}{dt}=-\lambda q(N,t)+\lambda\sum_{s=0}^{N-2}\frac{(\lambda\tau)^s}{s!}e^{-\lambda\tau}q(N-1-s,t-\tau)
\end{equation}

The gain term reflects the fact that blockage only occurs with $N$ particles, $N-1$ terms correspond to the cases where there 
may be from $0$ to $N-1$ particles in the channel. 

For $n>N$,  the dynamics of $q(n,t)$ for $n >N$ can be approximated as follows
\begin{align}\label{eq:m2}
&\frac{d q(n,t)}{dt}=-\lambda q(n,t)+\lambda q(n-1,t-\tau)e^{-\lambda\tau}\nonumber\\
&+\lambda \sum_{s=1}^{N-2}\int_0^\tau dt_1 K_s(t_1)e^{-\lambda\tau}q_s(n-1-s,t-\tau-t_1)
\end{align}
where we have introduced a kernel $K_s(t)$.
We then consider two physical situations. In the first, the last $s$  particles are assumed to have entered the channel in an infinitesimal 
time interval and the kernel is given by 
$K_s(t)=\frac{(\lambda\tau)^s}{s!}\delta(t)$.
This choice overestimates the survival probability. $N-2$ particles can be in the channel (so can enter between $t-\tau$ and $t$) when a new particle enters. 
The other particles enter between time $0$ and $t-\tau$. This fails to take into account some blocking.
In the second case we take $K_s(t)=\lambda \frac{(\lambda t)^{s-1}}{(s-1)!} e^{-\lambda t}$ which is proportional to the 
probability that $s-1$ particles enter in $(0,t)$. This choice underestimates the survival probability.
 When a particle enters at time $t$ there may be a maximum of $N-2$ particles in the channel to avoid blocking. 
If a particle arrives at a time $t_1$ between $t-\tau$ and $t$, there may be a maximum of $N-3$ particles between $t$ and $t-t_1$ 
and no particle between $t-t_1$ and $t-t_1-\tau$.

Taking the  Laplace transform of Eq.(\ref{eq:m2}),  we calculate two different generating functions corresponding to the two kernels,  
and the corresponding mean survival times.
These bracket the exact value and for $\lambda \tau\ll 1$ the two solutions approach the same limit:
\begin{equation}\label{eq:tN}
\langle t \rangle=\frac{(N-1)!}{(\lambda \tau)^N}
\end{equation}

To obtain exact results for $N\ge 4$ is a challenging problem.
We therefore finish this section by presenting some numerical results that illustrate the general trends. 
The inset of  Fig. \ref{fig:tbarN} compares the asymptotic behavior for mean blocking time, Eq.(\ref{eq:tN}) 
with simulation results
for $N=2$ to $N=5$. We observe that the scaling law  provides 
provides a good description of the process for $\lambda\tau \leq 0.5$.

\begin{figure}[t]
\begin{center}
\includegraphics[width=8cm]{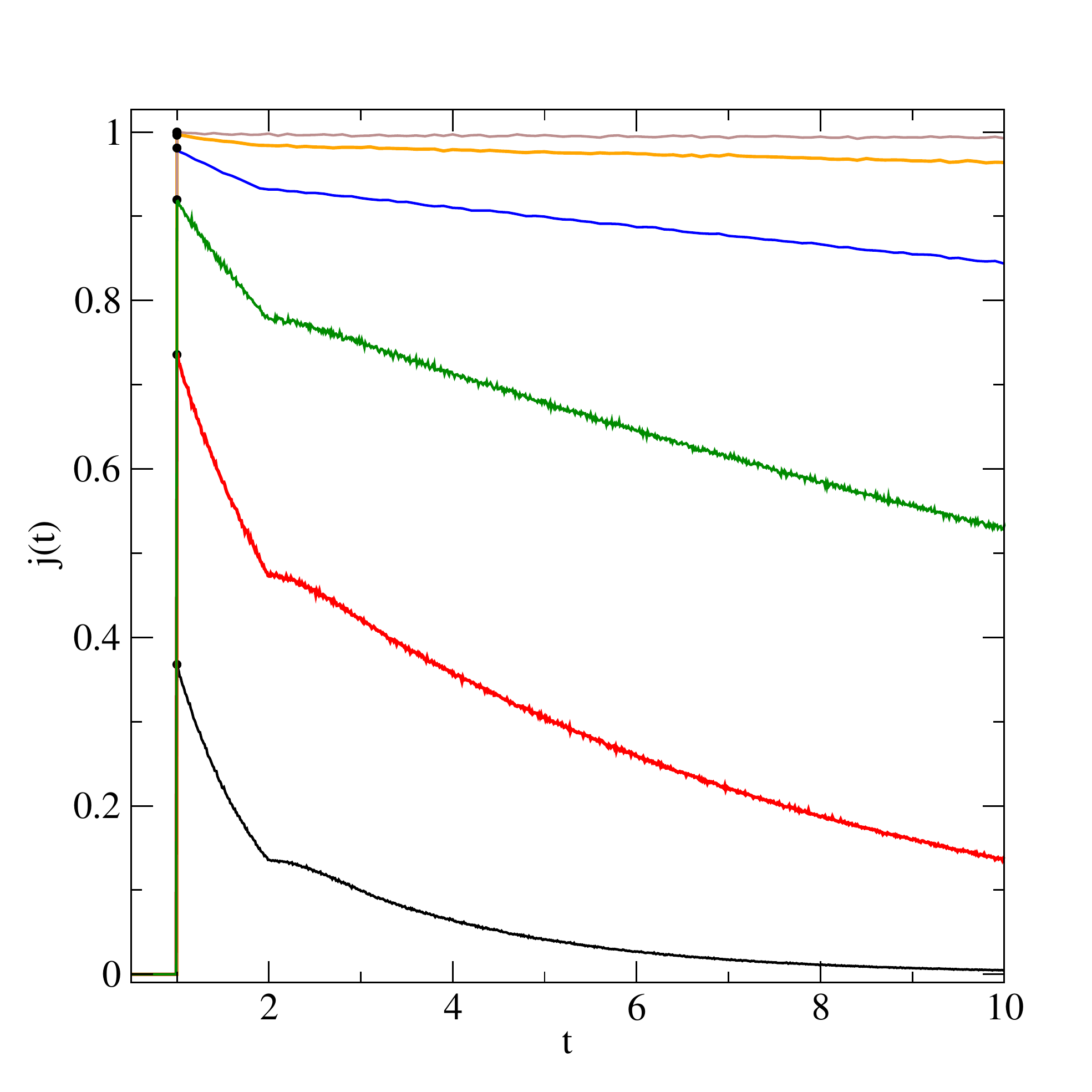}\\
\includegraphics[width=8cm]{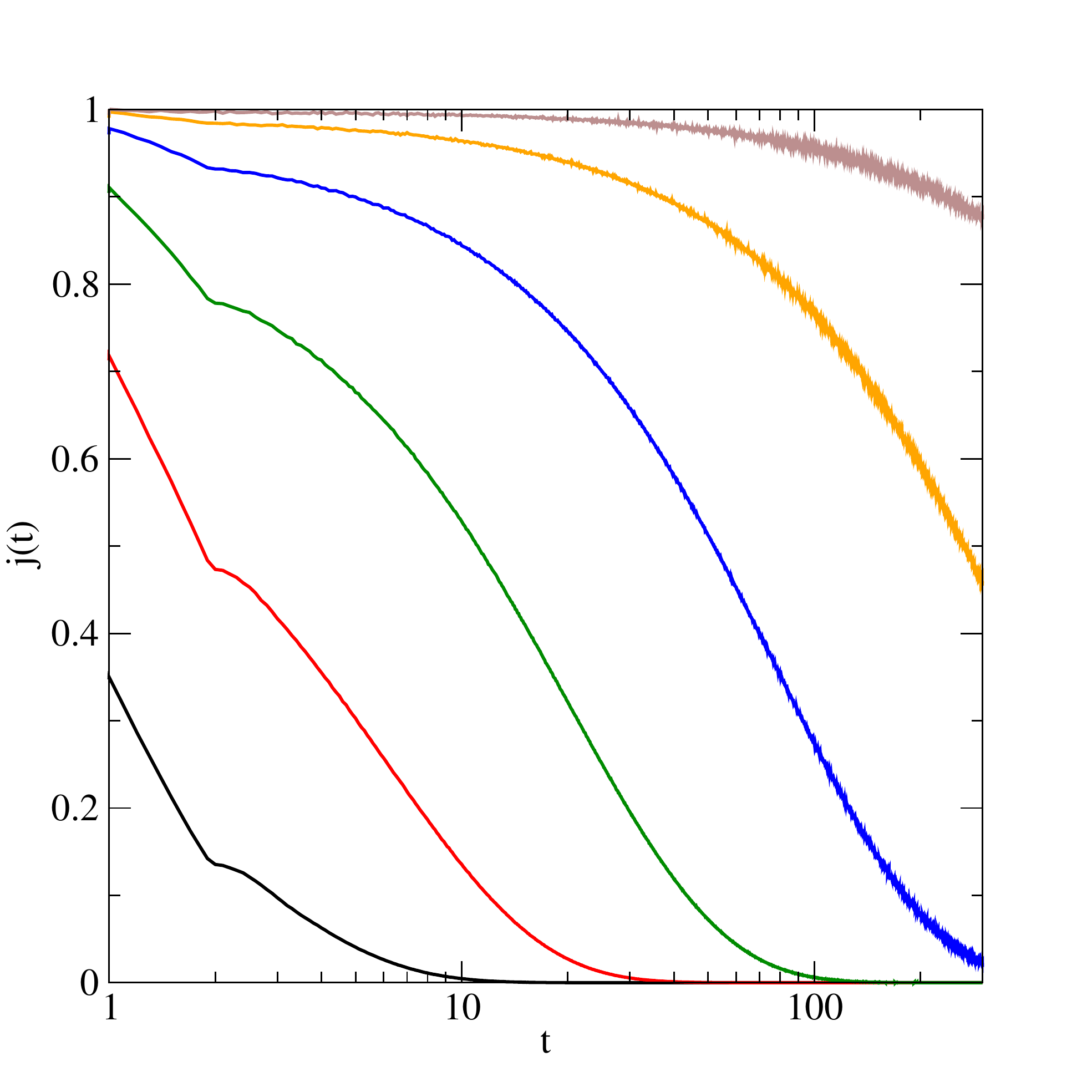}
\end{center}
\caption{Mean flux $j(t)$ versus time $t$  for $N=2,3,4,5,6,7$ (from bottom to top) for a Poisson distribution with $\lambda=1$. Top: short time behavior. 
At $t=1$,  circles correspond to the exact values of the mean flux, Eq.(\ref{eq:fluxexactN}). 
Bottom: linear-log plot showing the long-time behavior. ($\tau=1$)}
\label{fig:flux}
\end{figure}

In Fig. \ref{fig:flux} we present numerical results for the mean  flux of exiting particles as 
a function of time. This quantity acquires a non-zero, maximum, value at  $t=\tau$ given by 
\begin{equation}
j(\tau)=\lambda\sum_{i=1}^{N-2}\frac{(\lambda\tau)^i}{i!}e^{-\lambda\tau}\label{eq:fluxexactN}
\end{equation}
This expression
corresponds to events where a particle exits between $t$ and $t+dt$ such that $0,1..,N-2$ particles are still in the channel. For $t>\tau$, 
we observe a drastic increase  of the characteristic decay time  as $N$ increases (see the lower figure of Fig. \ref{fig:flux}). 
For $N=2$, $j(t)$ is very 
small for $t>3\tau$, while for $N=7$, the flux is almost constant during 
two decades.

\section{Correlations}\label{sec:cor}
We now consider the time correlation function $C(t)$ that represents the  density function that any two particles have a time separation $t$.
$C(t)$ can be expressed as the sum of  partial correlation functions $c(n,t)$ 
that correspond to the probability density that the first and last particles of sequence of $n+1$ particles
are separated by $t$.
\begin{equation}
 C(t)=\sum_{n=1}^{\infty}c(n,t)\label{eq:correldef}
\end{equation}
The partial correlation function $c(n,t)$, the joint probability of having 
a particle at $t=0$ and the $n$th particle at time $t$,  can be written as
\begin{align}
 c(n,t)&= \int_0^{\infty}\prod_{i=1}^{N-2+n} dt_i c^{(N-2)}(t_1,...,t_{N-2}) \prod_{j=N-1}^{N-2+n} \psi(t_j)\nonumber\\
 &\times\left[\prod_{j=1}^{n}\theta\left(\sum_{m=0}^{N-2}t_{j+m}-\tau\right)\right]\delta\left(t-\sum_{i=N-1}^{N+n-2}t_i\right) \label{eq:correldef2}
\end{align}
where $c^{(N-2)}(t_1,...,t_{N-2})$ is the joint probability of having $N-1$ particles such that the first and the second particles are separated 
by a duration of $t_1$, the second and the third particles by a duration of $t_2$,..
and the $N-2$ and $N-1$ particles by $t_{N-2}$. We can write this probability as
\begin{align}
 c^{(N-2)}(t_1,...,t_{N-2})=&\int dt_0  c^{(N-2)}(t_0,...,t_{N-3})\nonumber \\
 &\times \psi(t_{N-2})\theta(\sum_{j=1}^{N-2}t_j-\tau)\label{eq:correldef3}
\end{align}

This definition of the correlation function considers  
all trajectories, including those that end before a given time $t$. As a result, the correlation function approaches zero at long time. 
It seems more interesting to keep only trajectories which have survived until at time $t$.

To generate a infinite sequence of particles corresponding to a trajectory of the model, let us consider the following rejection-free algorithm: 
Accounting for  the constraints of the model (only less than $N$ particles must enter the channel in the duration of time $\tau$) without interrupting the traffic, one introduce the discrete stochastic equation
\begin{equation}\label{eq:stochasticdiscrete}
 t_n=\max(\tau-\sum_{j=1}^{N-2}t_{n-j},0)+\eta
\end{equation}
where $\eta$ is a random number generated from the $\psi$ distribution and $t_{n-j},j=1,N-2$ are the time intervals of
between the $N-2$ previously entering particles.

In order to compute the correlation function associated with this rejection-free algorithm, we replace  
$\psi(t_i)$ in Eqs.(\ref{eq:correldef2},\ref{eq:correldef3}) with 
\begin{equation}\label{eq:replacepsi}
\psi(t_i-\max(\tau-\sum_{j=1}^{N-2}t_{i-j},0)). 
\end{equation}

The partial correlation function $c(n,t)$ can be also expressed as the average over the event of having a first and $n+1$th particles separated by 
a time duration $t$
\begin{equation}
 c(n,t)=\langle\delta (t-\sum_{i=1}^n t_i) \rangle
\end{equation}

The conservation of the probability reads 
\begin{equation}
 \int_0^{\infty} dt c(n,t)=1
\end{equation}

By summing over $n$, the integral correlation function $C(t)$ is given at long time by
\begin{equation}
 \int_0^t dt' C(t')=\langle n(t) \rangle 
\end{equation}
where $\langle n(t) \rangle $ is the mean number of particles along a trajectory for a time duration  $t$. At large $t$, 
this quantity goes to a constant because we only consider
trajectories that have survived.
By using that $C(t)$ goes to a constant at long time (due to to the decay of the memory between particles that entered with a
large time difference) ($C(t) \rightarrow C_{\infty}$),
we infer that $C_{\infty}=1/ \bar{t}$ where $\bar{t}$ is the average separation in time between successive particles.

We now focus on $N=2$ and $N=3$ by using the rejection-free trajectories for which  exact solutions can be obtained.

\subsection{$N=2$}
The partial correlation function $c(n,t)$ is simply given as the product of integrals on each independent interval.
Eq.(\ref{eq:stochasticdiscrete}) is very simple $t_n=\tau+\eta$, which means that $\psi(t)$ is replaced with $\psi(t-\tau)$ in Eq.(\ref{eq:replacepsi}).
Therefore, the  Laplace transform of $c(n,t)$ is  given by
\begin{align}\label{eq:cnu}
 \tilde{c}(n,u)&=\left(\int_\tau^\infty dt \psi(t-\tau) e^{-tu}\right)^n=\tilde{c}(1,u)^n
\end{align}
This results from the fact that  successive events are not correlated. 

Inserting Eq.(\ref{eq:cnu}) in Eq.(\ref{eq:correldef}) we obtain
\begin{equation}\label{eq:ctotu}
 \tilde{C}(u)=\frac{\tilde{c}(1,u)}{1-\tilde{c}(1,u)}
\end{equation}

At long time, $C(t)$ approaches a constant value
corresponding to a constant mean density. By using the factorization property, $\tilde{c}(n,u)=\tilde{c}(1,u)^n$, 
and the expansion $\tilde{c}(1,u)=\tilde{c}(1,0)+u\partial\tilde{c}(1,u)/\partial u|_{u=0}+O(u^2)$ one can show
that $C(\infty)=\lim_{u\rightarrow 0}u\tilde{C}(u)=1/\bar{t}$ where $\bar{t}=\int_0^{\infty}tc(1,t)dt=-\partial\tilde{c}(1,u)/\partial u|_{u=0}$ is the
average interval between particles. That is, the smaller the average separation in time between successive particles, the larger the
steady state value of the time correlation function.

For a Poisson distribution $\psi(t)=\lambda e^{-\lambda t}$ we find
\begin{equation}
 \tilde{C}(u)=\sum_{n=1}^{\infty}\left(\frac{\lambda}{\lambda+u}\right)^{n}e^{-n u\tau}\label{eq:correllapla}
\end{equation}
The inverse Laplace transform gives an explicit expression
\begin{equation}\label{eq:correl2}
C(t)=\lambda\sum_{n=1}^{\infty}\theta(\lambda(t-n\tau))\frac{(\lambda(t-n\tau))^{n-1}e^{-\lambda (t-n\tau)}}{(n-1)!}
\end{equation}

\begin{figure}[t]
 \begin{center}
 \includegraphics[angle=-90,width=9cm]{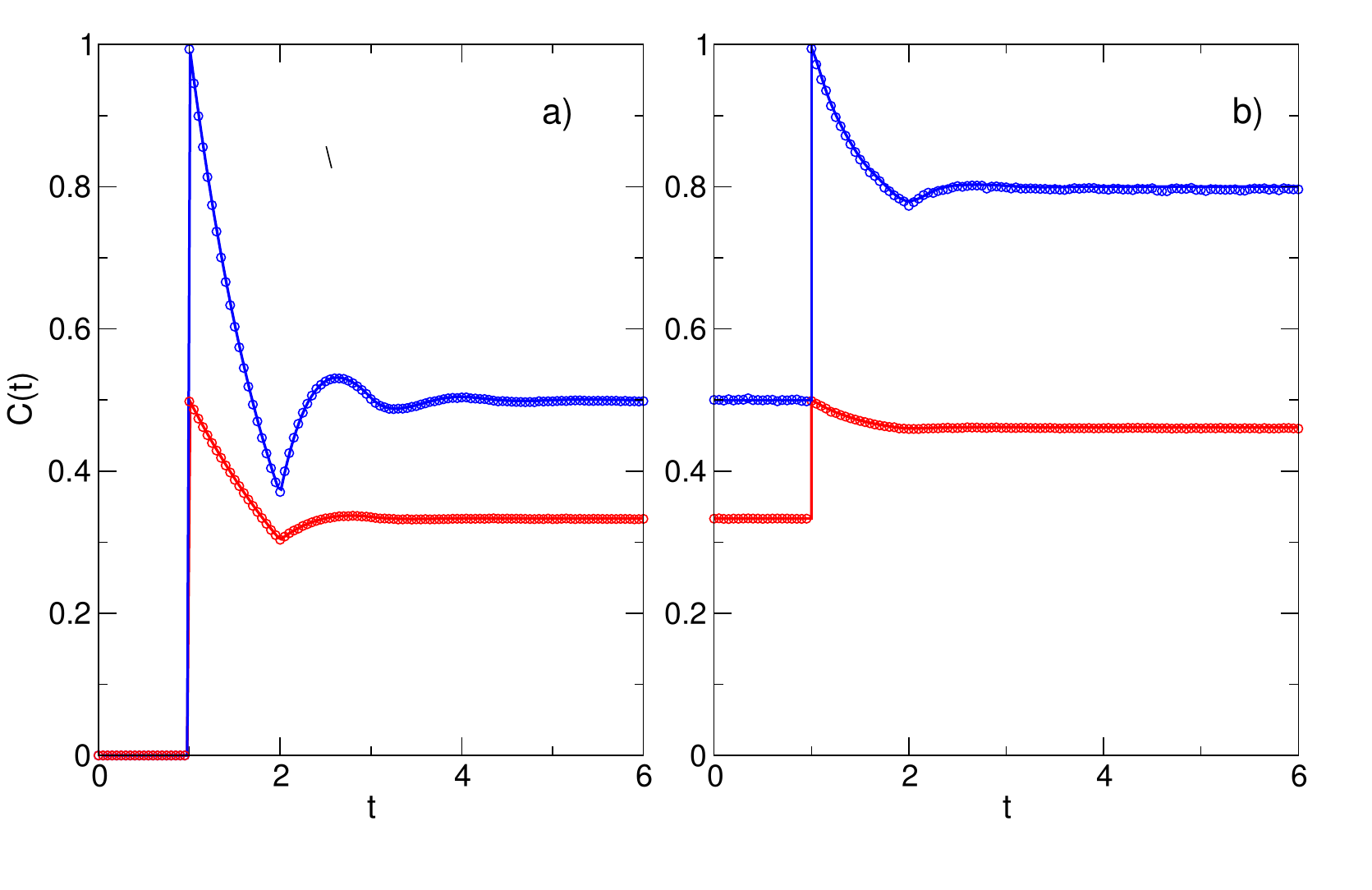}
 \end{center}
 \caption{Correlation functions for (a) $N=2$ and (b) $N=3$  for $\lambda=1$ and 0.5 (lower curves). 
 The solid lines correspond to Eq. (\ref{eq:correl2}) ($N=2$) and Eq. (\ref{eq:correl3}) ($N=3$) and circles to numerical simulations.}
 \label{fig:correN23}
\end{figure}

Figure \ref{fig:correN23}(a) shows $C(t)$ for two values of $\lambda\tau$. As expected, $C(t)$ is strictly equal to $0$
for $t < \tau$ since no particle can be inserted if the delay between two successive
particles is less than $\tau$. The maximum of $C(t)$ is obtained at $t=\tau$ where $C(\tau)=\lambda$ and decreases to $0$ at large $t$. 
Note that a cusp is present at $t=2\tau$, a similar  behavior  observed for the other quantities such as the
flux and the survival probability. 
In the long time limit $C(t=\infty)=\lim_{u\rightarrow 0}u\tilde{C}(u)=\lambda/(1+\lambda\tau)$.

 It is also interesting to note
that  correlation function Eq.(\ref{eq:correl2})  corresponds to  the density correlation function of the positions of the particle centers in 
a hard rod fluid of density $\rho$ with $\lambda=\rho/(1-\rho)$.

\subsection{$N=3$}

For $N= 3$, the discrete stochastic equation, Eq.(\ref{eq:stochasticdiscrete}), becomes 

\begin{equation} 
 t_n=\max(\tau-t_{n-1},0)+\eta
\label{eq:un}
\end{equation}
where  $t_n$ denotes the time interval between the $n-1$ and $n$ particles and $\eta$ is a random number
chosen with an exponential probability distribution $\lambda e^{-\lambda t}$.
In queuing theory  this equation is known as the Lindley-type equation\cite{Vlasiou2007,Boxma2007,Jodra2010}.

For the Poisson distribution $\psi(t)$,
Eqs.(\ref{eq:correldef2},\ref{eq:correldef3}) with Eq.(\ref{eq:replacepsi}) gives  
\begin{align}
c(n,t)=&\int_0^{\infty} dt_0 c(1,t_0)\delta(t-\sum_{i=1}^n t_i)\nonumber \\
& \prod_{i=1}^{n}(\int_{\max(\tau-t_{i-1},0)}^{\infty} dt_i \lambda e^{-\lambda(t_i -\max(\tau-t_{i-1},0)))}\label{eq:pn}
\end{align}
and 
\begin{align}
 c(1,t)&= \int_{Max(\tau-t,0)}^{\infty} dt_1  c(1,t_1)  \lambda e^{-\lambda(t-Max(\tau-t_1,0))}
 \label{eq:correl1}
\end{align}

Note that the constraint applies to two consecutive intervals, i.e the arrival time between $3$ consecutive particles is greater than $\tau$. 
Consequently, the partial 
correlation $\tilde{c}(n,u)$ is never the product of smaller correlation functions, as for the $N=2$ model.  

Because the kinetics were obtained exactly in the previous section only for the Poisson distribution, 
we restrict  our analysis to this distribution.

From Eq.(\ref{eq:correl1}), one easily shows that $c(1,t)$ is constant for $t>\tau$. 
For $t<\tau$, by taking the derivative of Eq.(\ref{eq:correl1}), one obtains 
\begin{equation} \frac{dc(1,t)}{dt}=\lambda (-c(1,t)+\theta (\tau-t)c(1,\tau-t))
\end{equation}
whose solution is 
\begin{equation}
 c(1,t)=\frac{\lambda }{1+\lambda \tau}(\theta (\tau-t)+e^{-\lambda (t-\tau)}\theta (t-\tau))\label{eq:p1}
\end{equation}

One can easily obtain the average time between two consecutively  particles in a trajectory.
\begin{align}\label{eq:tbarN3}
 \bar{t}&=\int_0^{\infty}dt c(1,t) t =\frac{(\lambda\tau+1)^2+1}{2\lambda(\lambda\tau+1)}
\end{align}
As might be expected, when the intensity $ \lambda\tau $ is high, the probability distribution is uniform within the first interval $[0,\tau]$ and equal to $\frac{1}{2}$. 
Conversely, when $ \lambda \tau$ tends to $ 0 $ the effect of 
the constraint is negligible, $ \bar{t}$ diverges as $ \frac {1} {\lambda} $, corresponding to the Poisson distribution.

It is easy to calculate the first few partial correlation functions by direct integration of Eq.(\ref{eq:pn}):  for instance, the probability $c(2,t)$ is given by
\begin{equation}
 c(2,t)=\frac{\lambda^2 t}{\lambda\tau+1} e^{-\lambda(t-\tau)} \theta(t-\tau)
\end{equation} 

To obtain a general expression of $c(n,t)$, we first take  the   Laplace transforms of Eq.(\ref{eq:pn})
\begin{align}
\tilde{c}(n,u)=&\int_0^{\infty} dt c(1,t)e^{-u t} m(n,t)
\end{align}
where $m(n,t)$ is auxiliary function  given by 
\begin{equation}
m(n,t)=\int_{\max(\tau-t,0)}^{\infty} dt' \lambda e^{-((u+\lambda) t' -\lambda\max(\tau-t,0))}m(n-1,t')\label{eq:m}
\end{equation}
The initial  condition is obviously,  $m(1,t)=1$.

Let us introduce the  generating function $G_m(z,t,u)$ of the auxiliary functions $m(n,t)$
\begin{equation}
G_m(z,t,u)=\sum_{n=1}^{\infty}z^{n-1} m(n,t)\label{eq:gf}
\end{equation}
 Inserting Eq.(\ref{eq:m}) in Eq.(\ref{eq:gf}), we obtain 
 \begin{align}
 G_m(z,t,u)&=1+z\int_{\max(\tau-t,0)}^{\infty} dt' G_m(z,t',u)\nonumber\\
 &\lambda e^{-((u+\lambda) t' -\lambda\max(\tau-t,0))}\label{eq:gcor1}
\end{align}
For $t>\tau$ the generating function is constant, $G_m(z,t,y)=G_m(z,\tau,u)$. 
For $t<\tau$, by taking the two partial derivatives of the integral equation Eq.(\ref{eq:gcor1}), one obtains

\begin{align}
\frac{\partial^2 G_m(z,t,u)}{\partial^2 t}&= z \lambda u e^{-u(\tau-t)}\left(G_m(z,\tau-t,u)\right.\nonumber\\
&\left.+\frac{\partial G_m(z,\tau-t,u)}{\partial t}\right)-\lambda \frac{\partial G(z,t,u)}{\partial t}
\end{align}
Simplifying we obtain

\begin{align}
\frac{\partial^2 G_m(z,t,u)}{\partial t^2}=& u \frac{\partial G_m(z,t,u)}{\partial t} + (u\lambda+\lambda^2-(\lambda z)^2e^{-u\tau})\times\nonumber\\
 &G_m(z,t,u)-u\lambda -\lambda^2-\lambda^2 z e^{-u(\tau-t)}\label{eq:generatingcor}
\end{align}
with boundary conditions (from Eq.(\ref{eq:gcor1})).

\begin{align}
\left\{ 
\begin{array}{ll}
G_m(z,0,u)&=1 + z G_m(z,\tau,u)\frac{\lambda  e^{-u\tau}}{u+\lambda}\\
\left.\frac{\partial G_m(z,t,u)}{\partial t}\right|_{t=\tau}&=z\lambda  G_m(z,0,u)-\lambda[G_m(z,\tau,u)-1]
\end{array}
\right. \label{eq:bccor}
\end{align}

whose solution is given by 

\begin{align}
G_m(z,t,u)=&A_1(z,u) e^{s_1 t}+B_1(z,u)e^{s_2 t}\nonumber\\
&+\frac{(u\lambda+\lambda^2+\lambda^2ze^{-u(\tau-t)})}{u\lambda+\lambda^2-(\lambda z)^2e^{-u\tau}}
\end{align}

where $s_{1,2}$ are the roots of the characteristic equation 
\begin{equation}
 s_{1,2}=\frac{1}{2}(u\pm\sqrt{(u+2\lambda)^2-4z^2\lambda^2 e^{-u\tau}})
\end{equation}

Finally we have 
\begin{align}
 G_m(z,t,u)&=\left(A_1(z,u) e^{s_1 t}+B_1(z,u)e^{s_2 t}\right. \nonumber\\
 &\left.+\frac{(u\lambda+\lambda^2+\lambda^2ze^{-u(\tau-t)})}{u\lambda+\lambda^2-(\lambda z)^2e^{-u\tau}}\right)\theta(\tau-t)\nonumber\\
 &+G_m(z,\tau,u)\theta(t-\tau)\label{eq:gcor2}
\end{align}
where $A(z,u)$ and $B(z,u)$ are determined by the boundary conditions, Eq.(\ref{eq:bccor}).

Using $G_m(z,t,u)$ and Eq.(\ref{eq:correldef}) we obtain the Laplace transform of the correlation function.
\begin{align}
 \tilde{C}(u)=&\int_0^{\infty}dt c(1,t)G_m(1,t,u)e^{-ut}\nonumber\\
 =&\frac{\lambda}{1+\lambda\tau}\left(\int_0^{\tau}dt G_m(1,t,u)e^{-ut}+G_m(1,\tau,u)\frac{e^{-u\tau}}{u+\lambda}\right)\label{eq:cnu1}
\end{align}
By inserting Eq.(\ref{eq:gcor2}) in Eq.(\ref{eq:cnu1}) we obtain 

\begin{align}\label{eq:correl3}
 \tilde{C}(u)=&\frac{\lambda}{(u+\lambda)(1+\lambda\tau)}\left[A_1(1,u)\left(-\frac{e^{-s_2\tau}(\lambda+s_1)-u-\lambda}{s_2}\right)\right.\nonumber\\
 &+B_1(1,u)\left(-\frac{e^{-s_1\tau}(\lambda+s_2)-u-\lambda}{s_1}\right) \nonumber\\
 &\left.+\frac{(u+\lambda)^2+e^{-u\tau}\lambda(u^2\tau+\lambda(u\tau-1))}{u(u+\lambda-\lambda e^{-u\tau})}\right]
\end{align}

Figure \ref{fig:correN23}(b) displays the correlation function $C(t)$ for $N=3$ versus time (with $\tau=1$).
As expected for $t\leq \tau$, $C(t)$ is constant and is equal to $\frac{\lambda}{1+\lambda\tau}$,
 because $c(n,t)=0,n>1$, and $c(1,t)$ is given by Eq.(\ref{eq:p1}), 
which is constant and different from $0$ in this time interval.
One also observes a discontinuity at $t=\tau$ and a long time limit equal to $\frac{2 \lambda(1+\lambda \tau)}{2+2\lambda\tau+\lambda^2\tau^2}$. 
We verify that, as for $N=2$, 
this is equal to $1/\bar{t}$ with $\bar{t}$ given by Eq.(\ref{eq:tbarN3}).

Comparing the correlation functions for $N=2$ and $N=3$ for the same values of $\lambda$ we note that the steady state values are higher for
$N=3$ corresponding to a shorter time interval between particles in the steady state. The oscillations are more pronounced for $N=2$ due to the
greater constraint imposed by the channel for smaller $N$ and hence greater correlations. \\

\section{Discussion}

The results presented in this article generalize the blocking model studied by Gabrielli et al. \cite{Gabrielli2013,TGV2015}. 
In order to examine 
the situation in which blockage is triggered by the simultaneous presence of $N>2$ particles 
in the channel and where the particle ingress
follows a general 
distribution of entry times, we have introduced an integral representation of the $n$ particle survival probabilities. 
For $N=3$, we have presented 
exact solutions for the mean time to blockage, Eqs.(\ref{temps},\ref{temps2}), as well as the correlation functions,
fluxes and other functions, 
for particles entering 
according to a Poisson distribution. 
For $N\ge 4$ obtaining an exact solution appears to be very challenging, but we have analyzed the generic features 
of the model using numerical
simulation. We also showed analytically that 
the mean time to blockage for small intensity and arbitrary $N$ diverges as a power of $N$, Eq.(\ref{eq:tN}). The is the
result of the fact that as $N$ increases, 
the channel exerts a weaker constraint on the incoming stream and blocking
is less likely.

Future directions include the development of a multichannel model, which 
can be applicable to filtration phenomenon \cite{PhysRevLett.98.114502}, and to consider systems with diffusive motion that are relevant for transport
through biological or synthetic nanotubes \cite{berezhkovskii2002}.

P. V. acknowledges illuminating discussions with Raphael Voituriez on discrete stochastic equations. J.T and P.V. acknowledge support from Institute of Mathematical
Sciences, National University of Singapore where the work was completed.


%
\end{document}